\RequirePackage{fix-cm}
\documentclass[twocolumn,epjc3]{svjour3}  
\smartqed  
\RequirePackage{graphicx,slashed}
\RequirePackage{mathptmx,amsmath,amssymb}      
\RequirePackage{flushend}
\RequirePackage{float}
\RequirePackage[numbers,sort&compress]{natbib}
\RequirePackage[colorlinks,citecolor=blue,urlcolor=blue,linkcolor=blue]{hyperref}
\RequirePackage[caption=false]{subfig}
\journalname{Eur. Phys. J. A}
\begin{document}

\title{On the critical end point in a two-flavor linear sigma model coupled to quarks
}


\author{Alejandro Ayala\thanksref{addr1,addr2},
        L. A. Hern\'andez\thanksref{addr1,addr2}, 
        M. Loewe\thanksref{addr2,addr3,addr4}
        Juan Cristobal Rojas \thanksref{addr5} and 
        R. Zamora\thanksref{addr6,addr7} 
}



\institute{Instituto de Ciencias Nucleares, Universidad Nacional Aut\'onoma de M\'exico, Apartado Postal 70-543, CdMx 04510, Mexico.\label{addr1}
          \and
          Centre for Theoretical and Mathematical Physics, and Department of Physics, University of Cape Town, Rondebosch 7700, South Africa.\label{addr2}
          \and
          Instituto de F\'isica, Pontificia Universidad Cat\'olica de Chile, Casilla 306, Santiago 22, Chile.\label{addr3}
          \and
          Centro Cient\'ifico-Tecnol\'ogico de Valpara\'iso CCTVAL, Universidad T\'ecnica Federico Santa Mar\'ia, Casilla 110-V, Valapara\'iso, Chile.\label{addr4}
          \and
          Departamento de F\'isica, Universidad Cat\'olica del Norte, Casilla 1280, Antofagasta, Chile.\label{addr5}
          \and
          Instituto de Ciencias B\'asicas, Universidad Diego Portales, Casilla 298-V, Santiago, Chile.\label{addr6}
          \and
          Centro de Investigaci\'on y Desarrollo en Ciencias Aeroespaciales (CIDCA), Fuerza A\'erea de Chile,  Santiago, Chile.\label{addr7}
}

\date{Received: date / Accepted: date}

\maketitle

\begin{abstract}
We use the linear sigma model coupled to quarks to explore the location of the phase transition lines in the QCD phase diagram from the point of view of chiral symmetry restoration at high temperature and baryon chemical potential. We compute analytically the effective potential in the high- and low-temperature approximations up to sixth order, including the contribution of the ring diagrams to account for the plasma screening properties. We determine the model parameters, namely, the couplings and mass-parameter, from conditions valid at the first order phase transition at vanishing temperature and, using the Hagedorn limiting temperature concept applied to finite baryon density, for a critical baryochemical potential of order of the nucleon mass. We show that when using the set of parameters thus determined, the second order phase transition line (our proxy for the crossover transition) that starts at finite temperature and zero baryon chemical potential converges to the line of first order phase transitions that starts at zero temperature and finite baryon chemical potential to determine the critical end point to lie in the region $5.02<\mu_B^{\mbox{\tiny{CEP}}}/T_c<5.18$, $0.14<T^{\mbox{\tiny{CEP}}}/T_c<0.23$, where $T_c$ is the critical transition temperature at zero baryon chemical potential.
\end{abstract}

\section{Introduction}\label{I}

One of the emerging challenges in the study of  strongly interacting matter subject to extreme conditions is to determine the existence and location of a critical end point (CEP) in the temperature ($T$) vs. baryon chemical potential ($\mu_B$) phase diagram. Research in this area is at the forefront of modern science and is relevant for several areas in physics, since it combines ideas not only from nuclear and particle physics but also from statistical physics applied to relativistic systems. Recall that at low $T$ and $\mu_B$, chiral symmetry is spontaneously broken. However, at large $T$ or $\mu_B$, where the QCD interaction weakens due to asymptotic freedom and the quark condensate is largely reduced, chiral symmetry is restored. Lattice QCD (LQCD) calculations show that for 2+1 (2) light flavors, the transition along the $T$ axis is a rapid crossover at a pseudocritical temperature $T_c\simeq 150-156$ MeV~\cite{lattice, Bernard, Cheng, Borsanyi2, Aoki2, Aoki3, Bazavov3, Bazavov4, Bhattacharya},  ($T_c\simeq 160-184$ MeV)~\cite{Tc2f}. On the other hand, a large number of effective models have shown that a first order phase transition happens for small $T$ and large $\mu_B$~\cite{models-first, Barducci,Barducci2,Barducci3,Berges,Halasz,Scavenius,Antoniou2,Hatta}. Therefore, when a line of first order phase transitions that starts at $T=0$ at a finite $\mu_B$ turns into a crossover, when $\mu_B=0$ at a finite $T$, the line must end and a CEP should exist.

Effective models are useful proxies to help identify the main characteristics of the QCD phase diagram. While no single model can be used to describe the whole extent of the phase diagram, these can be used to explore different regions with varying degrees of sophistication and inclusion of effective degrees of freedom. For instance, in Ref.~\cite{Herbst} a Polyakov-quark-meson model with quantum fluctuations, is employed to map the deconfinement and chiral symmetry restoration transitions. An important finding of that work is that the crossover region for one and the other transition coincide within a band representing the width of the susceptibility peak and that the width of such band shrinks as $\mu_B$ increases up to the region where a CEP at low temperature values is found. Although other effective models find non-coincident CEPs for the deconfinement and chiral symmetry transitions~\cite{Benic}, given that at least one class of models do find coincident transition lines, it should be  possible to explore the phase diagram emphasizing independently either the deconfinement or the chiral aspects of the transition.

LQCD calculations cannot be easily extended to finite $\mu_B$ due to the severe sign problem~\cite{sign}. However, a Taylor expansion technique, valid for small $\mu_B$, can be used to estimate the curvature of the  pseudocritical transition line. From this approach, it has been shown that this line has a negative and small curvature~\cite{taylor0,taylor1,taylor2,taylor3,deForcrand:2008,Bonati:2018,Bazavov2018}. Another useful LQCD technique is the analytic continuation from imaginary to real $\mu_B$ to extrapolate thermodynamical 
observables~\cite{Borsanyi,Guenther2018,Kashiwa,Cea,Bonati,Bonati:2015,Bellwied,Cea:2015}. Using Taylor expansions it has been shown that the disfavored region for the CEP location corresponds to $\mu_B^{\mbox{\tiny{CEP}}}/T_c^{\mbox{\tiny{CEP}}}\lesssim 1.8$ for $T_c^{\mbox{\tiny{CEP}}}\simeq 135$ MeV~\cite{Bazavov2}.

Much effort has been devoted to pin down the CEP position, both from the theoretical and the experimental fronts. On the theory side, a large number of techniques has been employed, including the sum rules method, Schwinger-Dyson equations, holography, functional renormalization and effective models~\cite{values, Ayala-Dominguez, Xin, Fischer, Fischer2, Lu, Shi, Contrera, Cui, Datta, Knaute, Antoniou, Rougemont, RMF, zamora1, zamora2,Schaefer,Marczenko}. A wide spread for the CEP location has been obtained (for a recent summary of theoretically computed CEP positions see Ref.~\cite{RMF}). On the experimental side, STAR has conducted the Beam Energy Scan (BES)-I~\cite{BESI,Luo:2017faz,Luo:2015doi,Adamczyk:2017wsl} analyzing heavy-ion collisions in the energy range 62.4 GeV $>\sqrt{s_{NN}}>$ 7.7 GeV. Future experiments~\cite{BESII,FAIR,NICA} will soon start providing data with higher statistics and lower collision energies for a wide range of colliding systems to explore deeper into the phase diagram.

The study of chiral symmetry restoration is a subject pertaining to the realm of full-fleshed QCD. However, given the LQCD limitations at finite $\mu_B$, effective models that incorporate chiral symmetry and its spontaneous breaking have proven to be quite useful. Among these, the linear sigma model with quarks (LSMq) has been used in Refs.~\cite{RMF,SS,Ayala-NPB, Ayala-IJMPA} to provide qualitative insight to explore both the location and nature of the transition lines as well as the CEP position. In these works it has been hinted that quantitative accuracy can be achieved when the model parameters (the couplings and the mass parameter) are determined not from vacuum conditions but instead from conditions describing the energy scales involved during the chiral symmetry restoration transition at finite $T$ and $\mu_B$. In this work we report on the prediction for the CEP location using the LSMq when the model parameters are determined from a first order phase transition at $T=0$ and a critical value for the baryon chemical potential $\mu_B^c$. Based on the Hagedorn limiting temperature concept applied to the case of finite chemical potential~\cite{Hagedorn,Satz}, we consider that the transition along the $T=0$ axis happens for $\mu_B^c\simeq m_N$, where $m_N\simeq 940$ MeV is the nucleon mass, and take this as the hadronic energy scale beyond where the thermodynamical description based on a resonance gas in thermal equilibrium does not hold anymore. The work is organized as follows: In Sec.~\ref{II} we introduce the LSMq. We compute the low and high-temperature approximations for the effective potential and the corresponding meson self-energies that are needed to describe the plasma screening properties. In Sec.~\ref{III} we determine the model parameters from the conditions that describe a crossover transition at finite $T$ and $\mu_B=0$ and a first order phase transition at finite $\mu_B$ and $T=0$. We finally conclude in Sec.~\ref{IV}. We leave for the appendices the details of the calculation leading to the expressions for the effective potential and the meson self-energies, as well as the solution to the equations that fix the model parameters and those to determine the vacuum stability.

\section{The linear sigma model with quarks}\label{II}

The Lagrangian for the linear sigma model when the two lightest quarks are included is given by
\begin{eqnarray}
   \mathcal{L}&=&\frac{1}{2}(\partial_\mu \sigma)^2  + \frac{1}{2}(\partial_\mu \vec{\pi})^2 + \frac{a^2}{2} (\sigma^2 + \vec{\pi}^2) - \frac{\lambda}{4} (\sigma^2 + \vec{\pi}^2)^2 \nonumber \\
   &+& i \bar{\psi} \gamma^\mu \partial_\mu \psi -g\bar{\psi} (\sigma + i \gamma_5 \vec{\tau} \cdot \vec{\pi} )\psi.
\label{lagrangian}
\end{eqnarray}

After spontaneous symmetry breaking, the Lagrangian for the LSMq is given by
\begin{align}
   \mathcal{L}&= \bar{\psi}(i\gamma^\mu \partial_\mu-M_q)\psi-g\bar{\psi}
   (\sigma+i\gamma_5\vec{\tau}\cdot \vec{\pi})\psi 
   - \frac{1}{2}M_\sigma^2\sigma^2\nonumber\\
   & - \frac{1}{2}M_\pi^2(\vec{\pi})^2
   -\frac{1}{2}(\partial_\mu\sigma)^2-
   \frac{1}{2}(\partial_\mu\vec{\pi})^2\nonumber\\
   &-\lambda v(\sigma^3+\sigma\vec{\pi}^2) -\frac{1}{4}\lambda (\sigma^4+2
   \sigma^2\vec{\pi}^2+\vec{\pi}^4)\nonumber\\
   & +\frac{a^2}{2}v^2-\frac{\lambda}{4}v^4 + \frac{m_\pi^2}{2} v(\sigma+v),
 \label{lagshift}
\end{align}
where $\psi$ is an $SU(2)$ isospin doublet of light quark flavors $(u,d)$, $\vec{\pi}=(\pi_1,\pi_2,\pi_3)$ is an isospin pion triplet and $\sigma$ is an isospin singlet. The $\sigma$-field has been shifted according to $\sigma\to\sigma + v$, where $v$ becomes the order parameter of the theory. As a consequence of the symmetry breaking the quarks, sigma and the three pions acquire dynamical masses
\begin{align}
 M_q&=gv, \nonumber \\
 M_\sigma^2&=3\lambda v^2-a^2, \nonumber \\
 M_\pi^2&=\lambda v^2-a^2,
 \label{masses}
\end{align}
respectively. Also, in order to consider a non-vanishing pion mass, we add to the Lagrangian an explicit symmetry breaking term and thus
\begin{eqnarray}
 \mathcal{L}\to\mathcal{L}'=\mathcal{L}+\frac{m_\pi^2}{2} v(\sigma+v),
\end{eqnarray}
where the vacuum pion mass is $m_\pi\simeq 139$ MeV. As a consequence of the explicit symmetry breaking, the effective potential at tree-level has a minimum at a value of $v$ given by 
\begin{eqnarray}
    v_0=\sqrt{\frac{a^2+m_\pi^2}{\lambda}}.
\end{eqnarray}
The coupling constants $\lambda$ and $g$,as well as the mass parameter $a$ are to be determined from physical conditions valid at the phase transition, whereas the minimum of the effective potential represents the order parameter that evolves when the system approaches chiral symmetry restoration at finite $T$ and/or $\mu_B$ and vanishes in the symmetric phase.

We emphasize that, although the strangeness degrees of freedom play an important role for the description of chiral symmetry restoration~\cite{Tawfik}, it is also true that models based on chiral symmetry increase their predictive power when the quark masses are small. Here we trade the inclusion of strangeness degrees of freedom in favour of a chiral symmetric model, broken only by a small pion mass.

Our strategy consists on computing the loop corrections to the tree-level potential. The first correction is the one-loop contribution. This contains vacuum and matter pieces. The vacuum piece is $v$-dependent and when added to the three-level potential it shifts $v_0$ and thus changing the vacuum masses. In order to avoid such change and to maintain the tree-level mass values, we add counterterms $\delta a^2$ and $\delta\lambda$ requiring that $v_0$ and the vacuum sigma mass do not change. This treatment is equivalent to the so called cancellation of {\it tadpole} contributions. The one-loop matter corrections contain $T$ as well as $\mu_B$ contributions. The chemical potential dependence comes from the quark sector whose chemical potential $\mu_q$ is related to $\mu_B$ by $\mu_B=3\mu_q$. The one-loop matter contributions corresponds to the mean field approximation for the system's energy. However, it is well known that for theories containing bosons whose mass can vanish at finite temperature and density, such as is the case during a phase transition, it is important to include corrections that account for the plasma screening effects~\cite{D&J}. Plasma screening can be incorporated into the treatment by including the resummation of the {\it ring diagrams}~\cite{Kapusta,LeBellac}. Screening is provided by the boson self-energy $\Pi_b$ which we approximate by its one-loop expression.

The effective potential up to the ring diagrams contribution as well as the boson self-energy can be analytically computed in the low and the high-temperature expansions. In the former, one considers that the largest energy scale is provided by $\mu_q$, such that $M/\mu_q$, $T/\mu_q\ll 1$, where $M$ is any of the particle masses. In the latter, it is only necessary to consider that $M/T\ll 1$, regardless of the relation between $T$ and $\mu_q$. The explicit expression for the effective potential in the low-$T$ expansion is given by
\begin{eqnarray}
V_{\text{LT}}^{\text{eff}}(v)&=&-\frac{(a^2+m_\pi^2+\delta a^2)}{2}v^2+\frac{(\lambda+\delta \lambda)}{4}v^4\nonumber \\
&-&3\Bigg\{\frac{ (M_\pi^2+\Pi_b^{\text{LT}})^2}{64\pi^2}\left[\ln\Big(\frac{4\pi a^2}{M_\pi^2+\Pi_b^{\text{LT}}}\Big)+\frac{1}{2}-\gamma_E\right]\nonumber \\
&+&T\bigg(\frac{T\sqrt{M_\pi^2+\Pi_b^{\text{LT}}}}{2\pi}\bigg)^{3/2}\text{Li}_{5/2}\Big(e^{-\sqrt{M_\pi^2+\Pi_b^{\text{LT}}}/T}\Big)\Bigg\} \nonumber \\
&-&\Bigg\{\frac{(M_\sigma^2+\Pi_b^{\text{LT}})^2}{64\pi^2}\left[\ln\left(\frac{4\pi a^2}{M_\sigma^2+\Pi_b^{\text{LT}}}\right)+\frac{1}{2}-\gamma_E\right]\nonumber \\
&+&T\bigg(\frac{T\sqrt{M_\sigma^2+\Pi_b^{\text{LT}}}}{2\pi}\bigg)^{3/2}\text{Li}_{5/2}\Big(e^{-\sqrt{M_\sigma^2+\Pi_b^{\text{LT}}}/T}\Big)\Bigg\} \nonumber \\
&+& N_c N_f\left\{ \frac{M_q^4}{16\pi^2}\left[\ln \left( \frac{4\pi a^2}{\left(\mu_q+\sqrt{\mu_q^2-M_q^2}\right)^2}\,\, \right)\right.\right. \nonumber \\
&-&\left.\gamma_E+\frac{1}{2}\right]-\frac{\mu_q\sqrt{\mu_q^2-M_q^2}}{24\pi^2}(2\mu_q^2-5M_q^2)\nonumber\\
&-&\frac{T^2\mu_q}{6}\sqrt{\mu_q^2-M_q^2}-\frac{7\pi^2T^4\mu_q}{360}\frac{(2\mu_q^2-3M_q^2)}{(\mu_q^2-M_q^2)^{3/2}}\nonumber \\
&+&\frac{31\pi^4 \mu_q M_q^4 T^6}{1008(\mu_q^2-M_q^2)^{7/2}}\Bigg\},
\label{VLT}
\end{eqnarray}
where for the fermion matter contribution to the one-loop effective potential, we have included terms up to ${\mathcal{O}}(T)^6$ using the expansion technique described in Ref.~\cite{chilenos}. Also, we have adopted the MS regularization scheme using $ae^{-1/2}$ as the renormalization scale. In the high-$T$ expansion it is given by
\begin{eqnarray}
V_{\text{HT}}^{\text{eff}}(v)&=&-\frac{(a^2+m_\pi^2+\delta a^2)}{2}v^2+\frac{(\lambda+\delta \lambda)}{4}v^4\nonumber \\
&-&3\Bigg\{\frac{M_\pi^4}{64\pi^2}\bigg[\ln \Big(\frac{a^2}{4\pi T^2}\Big)+\frac{1}{2}-\gamma_E\bigg]+\frac{\pi^2 T^4}{90}\nonumber \\
&-&\frac{T^2 M_\pi^2}{24}+\frac{T(M_\pi^2+\Pi_b^{\text{HT}})^{3/2}}{12\pi}+\frac{\zeta(3)M_\pi^6}{96\pi^4 T^2}\Bigg \}\nonumber \\
&-&\Bigg\{ \frac{M_\sigma^4}{64\pi^2}\bigg[\ln \Big(\frac{a^2}{4\pi T^2}\Big)+\frac{1}{2}-\gamma_E\bigg]+\frac{\pi^2 T^4}{90}\nonumber \\
&-&\frac{T^2 M_\sigma^2}{24}+\frac{T(M_\sigma^2+\Pi_b^{\text{HT}})^{3/2}}{12\pi}+\frac{\zeta(3)M_\sigma^6}{96\pi^4 T^2} \Bigg\}\nonumber \\
&+&\frac{N_c N_f}{16\pi^2}\Bigg\{ M_q^4 \bigg[ \ln \Big(\frac{a^2}{4\pi T^2} \Big)+\frac{1}{2}-\gamma_E\nonumber \\
&-&\psi^{(0)}\Big(\frac{1}{2}+\frac{\mathrm{i}\mu_q}{2\pi T}\Big)-\psi^{(0)}\Big(\frac{1}{2}-\frac{\mathrm{i}\mu_q}{2\pi T}\Big) \bigg]\nonumber \\
&-&8M_q^2T^2\big[\text{Li}_2(-e^{\mu_q/T})+\text{Li}_2(-e^{-\mu_q/T})\big]\nonumber \\
&+&32T^4\big[\text{Li}_4(-e^{\mu_q/T})+\text{Li}_4(-e^{-\mu_q/T})\big]\nonumber \\
&+&\frac{M_q^6}{6T^2}\left[\psi^{(2)}\Big(\frac{3}{2}+\frac{\mathrm{i}\mu_q}{2\pi T}\Big)\right.\nonumber \\
&+&\left.\psi^{(2)}\Big(\frac{3}{2}-\frac{\mathrm{i}\mu_q}{2\pi T}\Big)\right]\Bigg\},
\label{VHT}
\end{eqnarray}
where for the matter contribution we have included terms up to ${\mathcal{O}}(M)^6$, using the expansion technique described in Ref.~\cite{D&J}, both from bosons and fermions.
The corresponding expressions for the boson self-energy in the low- and high-temperature approximations are given by \cite{RMF}
\begin{equation}
    \Pi_b^{\text{LT}}=N_c N_f g^2\bigg(\frac{\mu_q^2}{2\pi^2}+\frac{T^2}{6}\bigg)
    \label{PiLT}
\end{equation}
and
\begin{eqnarray}
    \Pi_b^{\text{HT}}&=&\frac{\lambda T^2}{2}-N_c N_f \frac{g^2 T^2}{\pi^2}\nonumber\\
    &\times&\left[\text{Li}_2(-e^{\mu_q/T})+\text{Li}_2(-e^{-\mu_q/T})\right],
    \label{PiHT}
\end{eqnarray}
respectively.

\section{The Phase diagram and the Critical end Point}\label{III}

In order to fix the model parameters $\lambda$, $g$ and $a$, we require that for $T=0$ and $\mu_q=\mu_B^c/3$, the phase transition is first order. This implies that the effective potential shows two degenerate minima, one at a finite value of $v$, say $\tilde{v}_0$ and the other at $v=0$. Notice that when the transition starts, the system is in a {\it mixed phase}. This means in particular that a fraction of the pions have a mass corresponding to the minimum at $\tilde{v}_0$ and the rest have a vanishing mass. We thus consider that the {\it effective} pion mass is the average between these two phases and that this mass could be computed as if the mass corresponds to an intermediate value of $v$, say $v_0'$, between $v=\tilde{v}_0$ and $v=0$. We chose $v_0'$ to correspond to the maximum of the barrier between the degenerate minima. This is equivalent to applying the so called {\it lever rule}~\cite{lever-rule}. Using this construction, we obtain two more conditions to determine $\tilde{v}_0$ and $v_0'$ that together with the conditions for the minima at $v=\tilde{v}_0$ and $v=0$ to be degenerate, form a system of four equations for our five unknowns, namely, $\lambda$, $g$, $a$, $\tilde{v}_0$ and $v_0'$. In order to solve the system we look for solutions fixing one of the parameters and choose $\lambda$ for this purpose. We thus find a family of solutions when varying $\lambda$. It turns out that the solutions exists only for a narrow window of $\lambda$ values.

With the found set of parameters we now proceed to explore the phase diagram. Starting from $T=0$ and using the low-temperature approximation for the effective potential, Eq.~(\ref{VLT}), we increase $T$ and find the values of $\mu_q$ for which the effective potential shows two degenerate minima, one at $v=0$ and a second one at a finite value of $v$. We continue with this procedure until when, for a certain pair of values ($\mu_q^{\mbox{\tiny{CEP\ LT}}},T ^{\mbox{\tiny{CEP\ LT}}}$), the effective potential stops producing first order phase transition. 

Next, we use the very same set of parameters to explore the phase diagram using the high-temperature approximation for the effective potential, Eq.~(\ref{VHT}). Starting from $\mu_q=0$, we find the temperature for which the effective potential shows a phase transition. Interestingly enough, this transition is second order 
and happens at a critical temperature $T_c\sim 180$ MeV that changes little when varying the value of $\lambda$, that in turn determines the values of the rest of the parameters. Notice that this critical temperature also corresponds to the maximum of the chiral susceptibility and as such it serves to identify the temperature that characterises the crossover transition temperature.

\begin{figure}[t!]
{\centering
\includegraphics[scale=0.58]{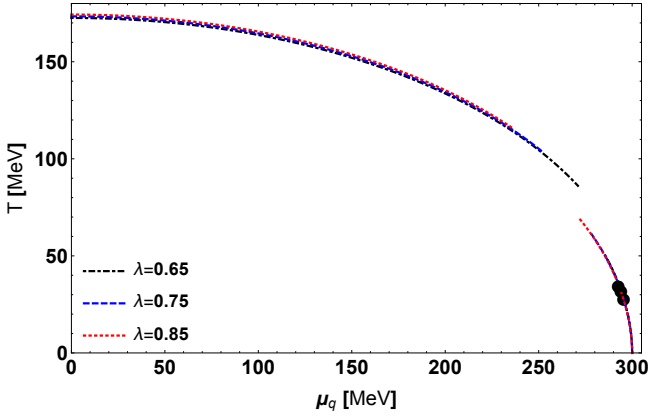}}
\caption{Phase diagram obtained from the high and low-temperature expressions for the effective potential for $\mu_q^c=300$ MeV and $\lambda=0.65,0.75,0.85$ that in turn correspond to $g=1.40,1.45, 1.50$ and $a=68.85, 82.05, 92.72$ MeV, respectively. The CEP is located within the full circles on each curve obtained using the low temperature expansion.}
\label{figura1}
\end{figure}
We continue with this procedure, varying the value of $T$ and finding the values of $\mu_q$ for which the effective potential still shows second order phase transitions until when, for a certain pair of values ($\mu_q^{\mbox{\tiny{CEP\ HT}}},T ^{\mbox{\tiny{CEP\ HT}}}$), the effective potential stops showing this kind of transitions. The CEP position is found to lie in the region $\mu_q^{\mbox{\tiny{CEP\ HT}}}<\mu_q^{\mbox{\tiny{CEP}}}<\mu_q^{\mbox{\tiny{CEP\ LT}}}$, $T^{\mbox{\tiny{CEP\ LT}}}<T^{\mbox{\tiny{CEP}}}<T^{\mbox{\tiny{CEP\ HT}}}$.

\begin{figure}[b!]
{\centering
\includegraphics[scale=0.58]{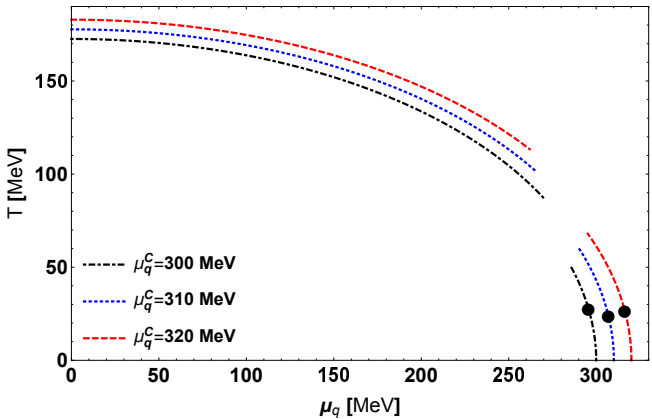}}
\caption{Phase diagram obtained from the high and low-temperature expressions for the effective potential for $\lambda=0.65$ with $\mu_q^c=300,310, 320$ MeV, that in turn correspond to $g=1.40,1.39,1.38$ and $a=68.85,80.00,90.56$ MeV, respectively. The CEP is located within the full circles on each curve obtained using the low temperature expansion.}
\label{figura2}
\end{figure}

Figure~\ref{figura1} shows the phase transition lines thus obtained for a fixed $\mu_q^c=300$ MeV and $\lambda=0.65,0.75,0.85$. Notice that the CEP appears in the region computed in the low-$T$ expansion.
Figure~\ref{figura2} shows the phase transition lines obtained when varying the value of $\mu_q^c$ between 300 and 320 MeV, for a fixed value of $\lambda$. Notice that changing $\lambda$ and $\mu_q^c$ produces a change in $T_c$ which nevertheless keeps being around $T_c\sim$ 180 MeV. We have checked that when using $\mu_B^c\simeq 1$ GeV ($\mu_q^c\simeq 333$ MeV), $T_c\gtrsim 190$ MeV, which is above the LQCD computed range 170 MeV $<T_c<$ 186 MeV~\cite{Tc2f}. We use this as a criterion to select $\mu_B^c\simeq m_N$.

\begin{figure}[t!]
{\centering
\includegraphics[scale=0.58]{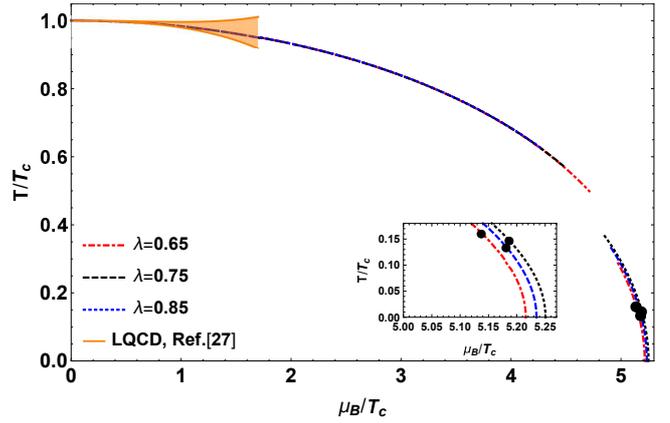}}
\caption{Phase transition lines obtained from the high and low-temperature approximations compared to the crossover transition region found by the LQCD Taylor expansion method in  Ref.~\cite{Bazavov2018}. The curves are plotted in terms of the scaled variables $\mu_B/T_c$ and $T/T_c$, where $T_c$ is the critical temperature at $\mu_B=0$ in each case. We have used the set of parameters $\mu_q^c=300$ MeV, $\lambda=0.65,0.75,0.85$ that in turn correspond to $g=1.40,1.45, 1.50$ and $a=68.85, 82.05, 92.72$ MeV, respectively. The CEP is located within the full circles on each curve obtained using the low temperature expansion.}
\label{figura3}
\end{figure}

Figure~\ref{figura3} shows the phase diagram in terms of the scaled variables $\mu_B/T_c$ and $T/T_c$ compared to the LQCD crossover transition region obtained from the Taylor series expansion method in Ref.~\cite{Bazavov2018}. For the figure we used $\mu_q^c=300$ MeV and varied the value of $\lambda$. We have checked that when using $\mu_q^c=310,\ 320$ MeV, the transition lines do not change appreciably. Notice that the theoretical  transition curves lie within the LQCD calculated region for low $\mu_B$ and that the CEP is located for rather low $T$ and high $\mu_B$.

\section{Conclusions}\label{IV}

In conclusion, we have used the LSMq to explore the QCD phase diagram from the point of view of the chiral symmetry restoration transition when increasing $T$ and $\mu_B$. For this purpose, we have used both a high- and low-tem\-per\-a\-ture analytical expansion of the effective potential up to sixth order including the contribution of the ring diagrams which encode the plasma screening properties. Based on the Hagedorn limiting temperature concept extended to finite chemical potential, and fixing $\mu_B^c\simeq m_N$, we have shown that the CEP position can be pinned down to lie in the region $5.02<\mu_B^{\mbox{\tiny{CEP}}}/T_c<5.18$, $0.14<T^{\mbox{\tiny{CEP}}}/T_c<0.23$, where $T_c$ is the critical transition temperature at $\mu_B=0$.

Notice that the region where the CEP was found corresponds to low-$T$, high-$\mu_B$ values where the experimental exploration becomes more challenging. These temperatures are also below those for the onset of color superconductivity and one could think that effects that are important to describe the latter should also be included in our formalism. Nevertheless, notice that color superconductivity is a phenomenon driven by density rather than by temperature in quark matter and pertains to the region where $\mu_B > m_N$, which lies outside the region of applicability of our approach. Nevertheless, it could be interesting to try including the contribution of other condensates appropriate for the description of color superconductivity. Also interesting is to further test the validity of the current approach, exploring the consequences for systems with an isospin imbalance. Work in this last direction is currently being pursued and will be reported elsewhere. 

\section*{ACKNOWLEDGEMENTS}

A. A. thanks the hospitality of M. L. at Instituto de F\'isica, PUC and of R. Z. at CIDCA-FACH during a research visit where this work was conceived. Support for this work has been received in part by UNAM-DGAPA-PAPIIT grant number AG100219 and by Consejo Nacional de Ciencia y Tecnolog\'ia grant number 256494. L. A. H. acknowledges support from a PAPIIT- DGAPA-UNAM fellowship. R. Z. would like to thank support from CONICYT FONDECYT Inicia\-ci\'on under grant No. 11160234. M. L. and J. C. R. acknowledge support from FONDECYT (Chile) under grants No. 1170107 and No. 1190192 and in addition, M. L. acknowledges support from CONICYT PIA/BASAL (Chile) grant No. 1190361.

\appendix

\section{Low temperature approximation}\label{AppeB}

Here, we show the explicit computation of the effective potential at low temperature (Eq.~(\ref{VLT})). We call $V_{\text{LT}}^{\text{b}}$ the boson piece and $ V_{\text{LT}}^{\text{f}} $ the fermion piece. First we compute the boson piece. Let us start from the one-loop
contribution
\begin{equation}
    V_{\text{}}^{\text{b}}=T\sum_n\int\frac{d^3k}{(2\pi)^3} \ln D(\omega_n,\vec{k})^{1/2},
    \label{1loopboson}
\end{equation}
where
\begin{equation}
D(\omega_n,\vec{k})=\frac{1}{\omega_n^2+k^2+m_b^2},
\end{equation} 
is the Matsubara boson propagator with $m_b^2$ being the square of the boson mass and $\omega_n=2n\pi T$ the Matsubara frequencies for boson fields. Calculating the sums over Matsubara frequencies, we obtain
\begin{eqnarray}
V_{\text{}}^{\text{b}}&=&\frac{1}{2\pi^2}\int_0^\infty dkk^2\biggl( \frac{\sqrt{k^2+m_b^2}}{2} \nonumber\\
&+& T \ln(1-e^{-\sqrt{k^2+m_b^2}/T})\biggr)\nonumber \\
&\equiv& V_{\text{}_{\text{vac}}}^{\text{b}}+ V_{\text{}_{\text{matt}}}^{\text{b}},
\end{eqnarray}
where we have separated the vacuum (vac) and matter (matt) contribution. First we calculate the vacuum contribution, 
\begin{eqnarray}
V_{\text{}_{\text{vac}}}^{\text{b}}=\frac{1}{2}\int \frac{d^3k}{(2 \pi)^3} \sqrt{k^2+m_b^2}, 
\end{eqnarray}
In order to carry out the calculation, we employ dimensional regularization. Using the well known expression
\begin{equation}
	\int \frac{d^Dk}{(2\pi)^D} \frac{1}{(k^2-m_b^2)^n}=\text{i}(-1)^n\frac{(m^2)^{2-\epsilon-n}}{(4\pi)^{2-\epsilon}}\frac{\Gamma(n-2+\epsilon)}{\Gamma(n)},
\end{equation}
with $D=d-2\epsilon $, this contribution can be written as
\begin{equation}
	V_{\text{}_{\text{vac}}}^{\text{b}}=\frac{\tilde{\mu}^{3-d}}{2}\int \frac{d^dk}{(2\pi)^d} \sqrt{k^2+m_b^2}.
    \label{vacuumbRD}
\end{equation}
In Eq.~(\ref{vacuumbRD}), we have explicitly $d=3$ and $n=-1/2$. Hence we have
\begin{equation}
	V_{\text{}_{\text{vac}}}^{\text{b}}=-\frac{m_b^4}{32\pi^2}\Gamma(\epsilon-2)\Big( \frac{4\pi \tilde{\mu}^2}{m_b^2} \Big)^\epsilon.
\end{equation}
Taking the limit $\epsilon \rightarrow 0$, we finally obtain
\begin{align}
	V_{\text{}_{\text{vac}}}^{\text{b}}=-\frac{m_b^4}{64\pi^2}\Big[ \ln \Big( \frac{4\pi \tilde{\mu}^2}{m_b^2}\Big)
    -\gamma_E+\frac{3}{2}+\frac{1}{\epsilon}\Big].
    \label{finalvaccumB}
\end{align}
We use the Minimal Subtraction scheme (MS). After fixing the renormalization scale to $\tilde{\mu}=a e^{-1/2}$, the final expression for the vacuum contribution is given by
\begin{equation}
V_{\text{}_{\text{vac}}}^{\text{b}}=-\frac{m_b^4}{64\pi^2}\Big[ \ln \Big( \frac{4\pi a^2}{m_b^2}\Big)-\gamma_E+\frac{1}{2}\Big]. \label{vaciob}
\end{equation}
Now, we calculate the matter contribution 
\begin{eqnarray}
V_{\text{}_{\text{matt}}}^{\text{b}}=\frac{T}{2\pi^2}\int_0^\infty dk \ln(1-e^{-\sqrt{k^2+m_b^2}/T})\label{b1},
\end{eqnarray}
when the temperature is low (LT) we can expand the logarithm in a Taylor series of the form
\begin{eqnarray}
\ln(1-e^{-\sqrt{k^2+m_b^2}/T})=-\sum_{n=1}^{\infty}\frac{e^{-n\sqrt{k^2+m_b^2}/T}}{n} \label{b2},
\end{eqnarray}
using Eq. (\ref{b2}) into Eq. (\ref{b1}), we obtain
\begin{eqnarray}
V_{\text{LT}_{\text{matt}}}^{\text{b}}=-\frac{1}{2\pi^2}\int_0^\infty dk \sum_{n=1}^{\infty}\frac{T e^{-n\sqrt{k^2+m_b^2}/T}}{n}.
\end{eqnarray}
We can make the change of variable $\omega=\sqrt{k^2+m_b^2}$ to get
\begin{eqnarray}
V_{\text{LT}_{\text{matt}}}^{\text{b}}&=&-\frac{1}{2\pi^2}\int_{m_b}^\infty d\omega \sum_{n=1}^{\infty}\frac{T e^{-n \omega/T}}{n}\nonumber \\
&=&-\frac{1}{2\pi^2} \sum_{n=1}^{\infty} \left(\frac{T m_b}{n} \right)^2 \left(\frac{T}{m_b  n}\right)^{1/2} K_2 \left(\frac{m_b n }{T}\right),
\end{eqnarray}
where $K_2$ is a Modified Bessel function of the second kind. Now, using the asymptotic expansion
\begin{equation}
   \lim _{{z\to \infty }}K_{\alpha}(z)=\sqrt{\frac{\pi}{2 z}} e^{-z},
\end{equation}
we obtain
\begin{equation}
V_{\text{LT}_{\text{matt}}}^{\text{b}}=-\frac{\sqrt{\pi}}{2\pi^2\sqrt{2}}\sum_{n=1}^{\infty} \left(\frac{T m_b}{n} \right)^2 \left(\frac{T}{m_b  n}\right)^{1/2} e^{-m_b n/T}.
\end{equation}
Using that 
\begin{equation}
    \sum_{n=1}^{\infty}\frac{1}{n^{5/2}}e^{-m_b/T}=\text{Li}_{5/2}(e^{-m_b/T}),
\end{equation}
where $\text{Li}$ is the  polylogarithm function, we get
\begin{eqnarray}
V_{\text{LT}_{\text{matt}}}^{\text{b}}=- T \left(\frac{m_b T}{2\pi} \right)^{3/2}  \text{Li}_{5/2}(e^{-m_b/T}).
\end{eqnarray}
Finally, we obtain for the boson case
\begin{eqnarray}
V_{\text{LT}}^{\text{b}}&=&-\frac{m_b^4}{64\pi^2}\Big[ \ln \Big( \frac{4\pi a^2}{m_b^2}\Big)-\gamma_E+\frac{1}{2}\Big]. \nonumber \\
&-& T \left(\frac{m_b T}{2\pi} \right)^{3/2}  \text{Li}_{5/2}(e^{-m_b/T}).
\end{eqnarray}
Now, we proceed to calculate the fermion case. The general expression for the one-loop correction at finite temperature for a fermion with mass $m_f$ is given by
\begin{equation}
    V^{\text{f}}=-T\sum_n\int\frac{d^3k}{(2\pi)^3} \text{Tr}[\ln S(\tilde{\omega}_n-i\mu_q,\vec{k})^{-1}],
    \label{1loopfermion}
\end{equation}
where
\begin{equation}
S(\tilde{\omega}_n,\vec{k})=\frac{1}{\gamma_0 \tilde{\omega}_n+\slashed{k}+m_f},
\end{equation}
is the Matsubara fermion propagator, $\tilde{\omega}_n=(2n+1)\pi T$ are the Matsubara frequencies for fermion fields and $\mu_q$ is quark chemical potential. We calculate the sum over the Matsubara frequencies to obtain
\begin{eqnarray}
V_{\text{}}^{\text{f}}&=&-2 \int \frac{d^3k}{(2 \pi^3)}(\sqrt{k^2+m_f^2}\nonumber \\ 
&+& T \ln(1+e^{-(\sqrt{k^2+m_f^2}-\mu_q)/T}) \nonumber \\
&+& T \ln(1+e^{-(\sqrt{k^2+m_b^2}+\mu_q)/T})    ) \nonumber \\
&\equiv& V_{\text{}_{\text{vac}}}^{\text{f}}+ V_{\text{}_{\text{mattI}}}^{\text{f}}+ V_{\text{}_{\text{mattII}}}^{\text{f}},
\end{eqnarray}
where we have separated the vacuum (vac) and the matter (mattI and mattII) parts. First we calculate the vacuum part. We notice that the fermion case differs from the boson case by an overall factor $-4$. Therefore if we multiply Eq.~(\ref{vaciob}) by $-4$, we get the vacuum one-loop contribution from one fermion field
\begin{equation}
V^{\textrm{f}}_{\text{}_{\textrm{vac}}}=\frac{m_f^4}{16\pi^2}\Big[ \ln \Big( \frac{4\pi a^2}{m_f^2}\Big)-\gamma_E+\frac{1}{2}\Big].
\end{equation}
Now we calculate the matter contribution, which we separate into two parts
\begin{equation}
   V_{\text{}_{\text{mattI}}}^{\text{f}}
=-2 \int \frac{d^3k}{(2 \pi^3)}T \ln(1+e^{-(\sqrt{k^2+m_f^2}-\mu_q)/T}), 
\end{equation}
and
\begin{equation}
   V_{\text{}_{\text{mattII}}}^{\text{f}}
=-2 \int \frac{d^3k}{(2 \pi^3)}T \ln(1+e^{-(\sqrt{k^2+m_f^2}+\mu_q)/T}).
\end{equation}
We have to analyse two cases. For the case at low temperature (LT), the first case is when $\sqrt{k^2+m_f^2}>\mu_q$. In this case $V_{\text{}_{\text{mattI}}}^{\text{f}}\rightarrow 0$ and $V_{\text{}_{\text{mattII}}}^{\text{f}}\rightarrow 0$. The second case is when $\sqrt{k^2+m_f^2}<\mu_q$. In this case $V_{\text{}_{\text{mattII}}}^{\text{f}}\rightarrow 0$, and 
\begin{equation}
   V_{\text{LT}_{\text{mattI}}}^{\text{f}}
=2 \int \frac{d^3k}{(2 \pi^3)} (\sqrt{k^2+m_f^2}-\mu_q) \theta(\mu_q-\sqrt{k^2+m_f^2}), \end{equation}
where $\theta(x)$ is the Heaviside step function. We can make the change of variable $\omega=\sqrt{k^2+m_b^2}$ to obtain
\begin{eqnarray}
   V_{\text{LT}_{\text{mattI}}}^{\text{f}}
&=&\frac{1}{\pi^2} \int_{m_f}^{\mu_q} d\omega \omega \sqrt{\omega^2-m_f^2}\nonumber \\
&=&\frac{1}{24\pi^2} \biggl[ \mu_q (5 m_f^2-2\mu_q^2) \sqrt{\mu_q^2-m_f^2} \nonumber \\
&+& 3 m_f^4 \ln\left(\frac{m_f}{\mu_c+\sqrt{\mu_c^2-m_f^2}}\right)\biggr].
\end{eqnarray}
Finally, we obtain for the fermion case
\begin{eqnarray}
V_{\text{LT}}^{\text{f}}&=& \frac{m_f^4}{16\pi^2}\Biggl[\ln \left( \frac{4\pi a^2}{\left(\mu_q+\sqrt{\mu_q^2-m_f^2}\right)^2} \right) \nonumber \\
&-&\left.\gamma_E+\frac{1}{2}\right]-\frac{\mu_f\sqrt{\mu_q^2-m_f^2}}{24\pi^2}(2\mu_q^2-5m_q^2) \Biggr] \nonumber \\
&\equiv&V_0^{\text{f}}.
\end{eqnarray}
To calculate the next order at low temperature, we make a Taylor expansion around $T=0$ \cite{chilenos}
\begin{align}
    V_{\text{LT}}^{\text{f}}=&V_0^{\text{f}}(\mu_q+xT)\Big |_{T=0}\nonumber \\
    &+\frac{\pi^2 T^2}{6}\frac{\partial^2}{\partial (xT)^2}V_0^{\text{f}}(\mu_q+xT)\Big |_{T=0}\nonumber \\
    &+\frac{\pi^4 T^4}{360}\frac{\partial^4}{\partial (xT)^4}V_0^{\text{f}}(\mu_q+xT)\Big |_{T=0} \nonumber \\
    &+\frac{31\pi^6 T^6}{15120}\frac{\partial^4}{\partial (xT)^4}V_0^{\text{f}}(\mu_q+xT)\Big |_{T=0}+\cdots,
    \label{1loopFLT}
\end{align}
Thus, we get
\begin{eqnarray}
V_{\text{LT}}^{\text{f}}&=&\frac{m_q^4}{16\pi^2}\Biggl[\ln \left( \frac{4\pi a^2}{\left(\mu_q+\sqrt{\mu_q^2-m_q^2}\right)^2}\,\, \right) \nonumber \\
&-&\left.\gamma_E+\frac{1}{2}\right]-\frac{\mu_q\sqrt{\mu_q^2-m_q^2}}{24\pi^2}(2\mu_q^2-5m_q^2)\nonumber\\
&-&\frac{T^2\mu_q}{6}\sqrt{\mu_q^2-m_q^2}-\frac{7\pi^2T^4\mu_q}{360}\frac{(2\mu_q^2-3m_q^2)}{(\mu_q^2-m_q^2)^{3/2}}\nonumber \\
&+&\frac{31\pi^4 \mu_q m_q^4 T^6}{1008(\mu_q^2-M_q^2)^{7/2}}.
\end{eqnarray}

\section{High temperature approximation}\label{AppeA}

Here, we show the explicit computation for the effective potential at high temperature,  Eq.~(\ref{VHT}). It has two pieces, bosonic and fermionic contributions. The effective potential is computed beyond the mean field approximation, up to order ring diagrams, within the imaginary time formalism. The computation is performed in the same fashion as~\ref{AppeB}. We begin with the boson piece, the one loop contribution is
\begin{equation}
    V_{\text{}}^{\text{b}}=T\sum_n\int\frac{d^3k}{(2\pi)^3} \ln D(\omega_n,\vec{k})^{1/2},
    \label{1loopb}
\end{equation}
where $D(\omega_n,\vec{k})$ is the boson propagator and $\omega_n$ the boson Matsubara frequencies. After the sum over $n$ is performed, we get
\begin{eqnarray}
V_{\text{}}^{\text{b}}&=&\frac{1}{2\pi^2}\int_0^\infty dkk^2\biggl( \frac{\sqrt{k^2+m_b^2}}{2} \nonumber\\
&+& T \ln(1-e^{-\sqrt{k^2+m_b^2}/T})\biggr)\nonumber \\
&\equiv& V_{\text{}_{\text{vac}}}^{\text{b}}+ V_{\text{}_{\text{matt}}}^{\text{b}},
\end{eqnarray}
where the one-loop contribution is written as the sum of two terms, the vacuum and matter pieces. The vacuum piece was computed in~\ref{AppeB} with the result
\begin{equation}
V_{\text{}_{\text{vac}}}^{\text{b}}=-\frac{m_b^4}{64\pi^2}\Big[ \ln \Big( \frac{4\pi a^2}{m_b^2}\Big)-\gamma_E+\frac{1}{2}\Big].
\end{equation}
The matter contribution is now computed in the high temperature approximation, which means that $T$ is the hard scale. Also, we consider $m_b/T \ll 1$. Then the matter contribution is
\begin{eqnarray}
V_{\text{}_{\text{matt}}}^{\text{b}}=\frac{T}{2\pi^2}\int_0^\infty dk \ln(1-e^{-\sqrt{k^2+m_b^2}/T})\label{b33},
\end{eqnarray}
and after expand Eq.~(\ref{b33}) in terms of powers  of $m_b/T$, the first five terms of the series are given by
\begin{align}
	V_{\text{HT}_{\text{matt}}}^{\text{b}}=&
-\frac{m_b^4}{64\pi^2}\ln \Big( \frac{m_b^2}{(4\pi T)^2}\Big)-\frac{\pi^2 T^4}{90}\nonumber \\
&+\frac{m_b^2 T^2}{24}-\frac{m_b^3 T}{12\pi}-\frac{\zeta(3)m_b^6}{96\pi^4T^2}.
\end{align}
Therefore, the boson one-loop contribution becomes
\begin{align}
   V_{\text{HT}}^{\text{b}}=&-\frac{m_b^4}{64\pi^2}\Bigg[ \ln\Big({\frac{a^2}{4\pi T^2}\Big)}-\gamma_E+\frac{1}{2}\Bigg]
   -\frac{\pi^2 T^4}{90}\nonumber \\
   &+\frac{m_b^2 T^2}{24}-\frac{m_b^3 T}{12\pi}-\frac{\zeta(3)m_b^6}{96\pi^4T^2}.
   \label{bosHTfinal}
\end{align}

In order to go beyond the mean field approximation, we consider the plasma screening effects. These can be accounted for by means of the ring diagrams. Since we are working in the high temperature approximation, we notice that the lowest Matsubara frequency is the most dominant term. Therefore we do not need to compute the other modes and it becomes
\begin{align}
    V^{\text{Ring}}&=\frac{T}{2}\int\frac{d^3k}{(2\pi)^3}\ln (1+\Pi_b^{HT}D(\vec{k}))\nonumber \\
    &=\frac{T}{4\pi^2}\int dk \ k^2 \Big \{ \ln(k^2+m_b^2+\Pi_B^{HT})   -\ln(k^2+m_b^2) \Big\}.
    \label{rings2}    
\end{align}
From Eq.~(\ref{rings2}), we see that both integrands are almost the same except that one is modified by the self-energy and the other one is not. Thus, after integration, we obtain that the ring diagrams contribution is
\begin{equation}
    V^{\text{Ring}}=\frac{T}{12\pi}(m_b^3-(m_b^2+\Pi_b^{HT})^{3/2}).
    \label{finalrings}
\end{equation}

On the other hand, in the same way that we compute the one-loop boson piece, we now calculate the fermion one. Then, we begin with
\begin{equation}
    V^{\text{f}}=-T\sum_n\int\frac{d^3k}{(2\pi)^3} \text{Tr}[\ln S(\tilde{\omega}_n-i\mu_q,\vec{k})^{-1}],
    \label{1lfermion}
\end{equation}
where $S(\widetilde{\omega}_n,\vec{k})$ is the fermion propagator, $\widetilde{\omega}_n$ the fermion Matsubara frequencies and $\mu_q$ the quark chemical potential. When we perform the sum over $n$, we get
\begin{eqnarray}
V_{\text{}}^{\text{f}}&=&-2 \int \frac{d^3k}{(2 \pi^3)}(\sqrt{k^2+m_f^2}\nonumber \\ 
&+& T \ln(1+e^{-(\sqrt{k^2+m_f^2}-\mu_q)/T}) \nonumber \\
&+& T \ln(1+e^{-(\sqrt{k^2+m_b^2}+\mu_q)/T})    ) \nonumber \\
&\equiv& V_{\text{}_{\text{vac}}}^{\text{f}}+ V_{\text{}_{\text{mattI}}}^{\text{f}}+ V_{\text{}_{\text{mattII}}}^{\text{f}},
\label{fermionsum}
\end{eqnarray}

In Eq.~(\ref{fermionsum}), we have one piece from the vacuum contribution, it was computed in~\ref{AppeB} and the result is
\begin{equation}
V^{\textrm{f}}_{\text{}_{\textrm{vac}}}=\frac{m_f^4}{16\pi^2}\Big[ \ln \Big( \frac{4\pi a^2}{m_f^2}\Big)-\gamma_E+\frac{1}{2}\Big],
\end{equation}
and two matter pieces, one for the particle and other one for the antiparticle. In the high temperature approximation, where $m_f/T\ll 1$, we proceed in a fashion entirely analogous to the boson case and obtain
\begin{align}
	V_{\text{HT}_{\text{matt}}}^{\text{f}}=&
\frac{1}{16\pi^2}\Bigg\{m_f^4\Bigg[\ln \Big( \frac{m_f^2}{(4\pi T)^2}\Big)\nonumber \\
&-\psi^{(0)}\bigg(\frac{1}{2}+\frac{i\mu_q}{2\pi T}\bigg)-\psi^{(0)}\bigg(\frac{1}{2}-\frac{i\mu_q}{2\pi T}\bigg)\Bigg]\nonumber \\
&-8m_f^2T^2\big[\text{Li}_2(-e^{\mu_q/T})+\text{Li}_2(-e^{-\mu_q/T})\big]\nonumber \\
&+32T^4\big[\text{Li}_4(-e^{\mu_q/T})+\text{Li}_4(-e^{-\mu_q/T})\big]\nonumber \\
&+\frac{m_f^6}{6T^2}\left[\psi^{(2)}\bigg(\frac{3}{2}+\frac{\mathrm{i}\mu_q}{2\pi T}\bigg)+\psi^{(2)}\bigg(\frac{3}{2}-\frac{\mathrm{i}\mu_q}{2\pi T}\bigg)\right]\Bigg\}.
\end{align}
Therefore, the fermion one-loop contribution is
\begin{align}
	V_{\text{HT}}^{\text{f}}=&
\frac{1}{16\pi^2}\Bigg\{m_f^4\Bigg[\ln \Big( \frac{a^2}{4\pi T^2}\Big)-\gamma_E+\frac{1}{2}\nonumber \\
&-\psi^{(0)}\bigg(\frac{1}{2}+\frac{i\mu_q}{2\pi T}\bigg)-\psi^{(0)}\bigg(\frac{1}{2}-\frac{i\mu_q}{2\pi T}\bigg)\Bigg]\nonumber \\
&-8m_f^2T^2\big[\text{Li}_2(-e^{\mu_q/T})+\text{Li}_2(-e^{-\mu_q/T})\big]\nonumber \\
&+32T^4\big[\text{Li}_4(-e^{\mu_q/T})+\text{Li}_4(-e^{-\mu_q/T})\big]\nonumber \\
&+\frac{m_f^6}{6T^2}\left[\psi^{(2)}\bigg(\frac{3}{2}+\frac{\mathrm{i}\mu_q}{2\pi T}\bigg)+\psi^{(2)}\bigg(\frac{3}{2}-\frac{\mathrm{i}\mu_q}{2\pi T}\bigg)\right]\Bigg\}.
\label{ferm1HT}
\end{align}

With the Eqs.~(\ref{bosHTfinal}), (\ref{finalrings}) and~(\ref{ferm1HT}) at hand, we can write the effective potential up to the ring diagrams contribution in the high temperature approximation.

\section{Fixing the set of parameters}\label{AppeC}

Following the idea in Sec.~\ref{III}, the set of equations, to determine the free parameters in the model, is
\begin{eqnarray}
V^{\text{eff}}\Big |_{v=0,T=0, \mu_q=\mu_B^c/3}&=&V^{\text{eff}}\Big |_{v=v_0',T=0, \mu_q=\mu_B^c/3} \nonumber \\
\frac{\partial}{\partial v}V(v, \lambda, g)\Big |_{v=\tilde{v}_{0},T=0, \mu_q=\mu_B^c/3} &=&0 \nonumber \\
\frac{\partial}{\partial v}V(v, \lambda, g)\Big |_{v=v_0',T=0, \mu_q=\mu_B^c/3} &=&0
\nonumber \\
a&=&\sqrt{\frac{4}{3}\lambda (4 \tilde{v}^2_{0}-v_0'^2)}\label{c36},
\end{eqnarray}
Note the last of Eqs. (\ref{c36}) comes from the level rule, where $m_{\pi}|_{v=\tilde{v}_0}=\frac{1}{2} m_{\pi}|_{v=v_0'}$. Also, we use $\lambda$ as an input. Hence, we have four equations for four unknowns (a, g, $\tilde{v}_{0}$, $v_0'$) from where we get the solution. It is remarkable that this solution exists only for a narrow window of $\lambda$ values.

\section{Vacuum stability conditions}\label{AppeD}
The vacuum stability conditions are introduced to ensure that $v_0$ and the sigma-mass maintain their tree level values, even after including the vacuum pieces stemming from the one-loop corrections. These conditions are
\begin{align}\label{v02}
\frac{1}{2v}\frac{dV^{\text{vac}}}{dv} \Big|_{v=v_0}&=0,\nonumber \\
\frac{d^2V^{\text{vac}}}{dv^2} \Big|_{v=v_0}&=2a^2+3m_\pi^2,
\end{align}
where $V^{\text{vac}}$ is the one-loop vacuum piece of the effective potential. The solution for the counterterms $\delta a^2$ and $\delta \lambda$ is given by 
\begin{eqnarray}
    \delta a^2&=&\frac{m_\pi^2}{2}
    -\frac{1}{16\pi^2 \lambda}\left\{\frac{}{} 3\lambda^2(6 a^2+4 m_\pi^2) -8 g^4(a^2+m_\pi^2) \right.\nonumber \\
    &+&\left.3 a^2 \lambda ^2 \Big[\ln  \Big(\frac{m_\pi^2}{a^2}\Big)+ \ln\Big(\frac{ 2a^2+3m_\pi^2}{a^2}\Big)\Big]\right\},
\end{eqnarray}
\begin{eqnarray}
\delta \lambda    &=&\frac{\lambda}{2}\left(\frac{ m_\pi^2}{a^2+ m_\pi^2}\right)\nonumber\\
    &-&\frac{1}{16\pi^2} \left\{\frac{}{}-16  g^4+24\lambda^2\right. -8 g^4 \ln  \left(g^2 \frac{(a^2+m_\pi^2)}{a^2\lambda}\right)\nonumber \\
    &+&\left. 3 \lambda ^2\left[ \ln  \left(\frac{m_\pi^2}{a^2}\right)
    +3 \ln  \left(\frac{ 2a^2+3m_\pi^2}{a^2}\right)\right]\right\}.
\end{eqnarray}

\end{document}